# Testing baryon number conservation in braneworld models with cold neutrons

Michaël Sarrazin*

*Research Center in Physics of Matter and Radiation (PMR),*
*Department of Physics, University of Namur, 61 rue de Bruxelles, B-5000 Namur, Belgium*

**Abstract**

In the context of multi-brane Universe models, matter swapping between two braneworlds is allowed leading to a baryon number violation in our visible world. An experimental framework is described to test such a phenomenology with cold neutrons thanks to a neutron-shining-through-a-wall experiment.

**1. Introduction**

Braneworld is a concept emerging from high energy physics in some attempts to get a Great Unified Theory or a Theory of Everything (Rubakov and Shaposhnikov, 1983; Davies, et al., 2008). Along this line of thought, our observable Universe can be considered as a three-dimensional space sheet (a 3-brane) embedded in a larger spacetime with N > 4 dimensions (the bulk). Although many puzzling problems (hierarchy, dark matter, …) could find some answers in such a theoretical framework, the proof of the existence of extradimensions or other branes is still lacking. As a consequence, there is a real necessity to explore the new physics which results from such exotic hypothesis in order to conceive new experimental ways to test it. While colliders are often associated with the quest of a new physics (when looking for Kaluza-Klein states for instance), it is also known that some detectable effects could be observed at low energy (Petit and Sarrazin 2005; Berezhiani and Bento, 2006; Abel, et al., 2008). In this way, an interesting phenomenology remains in scenarios in which many branes coexist in the bulk (Petit and Sarrazin, 2005; Sarrazin and Petit, 2010 and 2012). Regarding the dynamics of spin-1/2 particles, it has been demonstrated that at low energies any two-brane Universe related to a domain wall approach is formally equivalent to a noncommutative two-sheeted spacetime $M_4 \times Z_2$ (Sarrazin and Petit, 2010) (see Fig. 1).

---

* *E-mail address:* michael.sarrazin@unamur.be

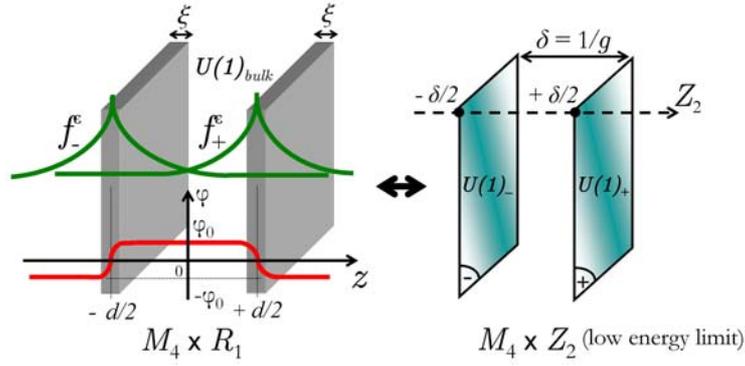

Fig. 1. A two-brane Universe made of two domain walls (two kink-like solitons) on a continuous $M_4 \times R_1$ manifold (on left), can be described by a noncommutative two-sheeted spacetime $M_4 \times Z_2$. φ is the scalar field which allows the existence of kink-like solutions (red curve), i.e. of domain walls by following the Rubakov-Shaposhnikov concept. The U(1) bulk gauge field is substituted by an effective $U(1)_+ \times U(1)_-$ gauge field in the $M_4 \times Z_2$ spacetime thanks to the Dvali–Gabadadze–Shifman mechanism which leads to the gauge field localization on the branes. Both branes can then be considered as separated by a phenomenological distance $\delta = 1/g$. $g$ is proportional to an overlap integral of the fermionic wave functions (green curves) of each 3-brane over the extra dimension.

From this approach, the following phenomenologies can be derived:
- Photon/hidden photon kinetic mixing
  Hidden photons could live in other branes. In a stringy context, a U(1) field on a hidden brane couples to the U(1) photon field of our brane through a one-loop process to give the photon-hidden photon kinetic mixing (Abel et al., 2004).
- Fermion/hidden fermion mass mixing
  Such a coupling is reminiscent of this considered in the mirror world approaches. The matter/mirror matter mixing is considered with neutron and mirror neutron for instance (Berezhiani and Bento, 2006).
- Fermion/hidden fermion geometrical mixing
  Such a coupling which involves the coupling constant g, should be also the most significant (Petit and Sarrazin, 2005; Sarrazin and Petit, 2010 and 2012).

Why two-brane physics is so interesting? Because it allows low-energy experiments to probe a new physics invisible at colliders. Roughly speaking brane thickness ξ is probed at colliders at energy E ∝ 1 / ξ. By contrast, in two-brane physics, the relevant parameter is the coupling $g$ which can be approximated for the purpose of our discussion by (Sarrazin and Petit, 2010):

$$g \approx (1/\xi)\exp(-kd/\xi) \tag{1}$$

with $d$ the real distance between each brane (see Fig. 1) and where $k \approx 1$ is a constant related to the fermion mass and the domain wall properties (Sarrazin and Petit, 2010). For branes at TeV scale, if $d \approx 50\,\xi$, for instance, then $g \approx 10^{-3}$ m$^{-1}$. This allows new physics at colliders, but also in cold neutrons facilities (such as ILL or ESS) (Sarrazin et al., 2012, Sarrazin and Petit, 2012). But now, what if we get branes at Planck scale? New physics is no longer accessible at colliders, but keeping $g \approx 10^{-3}$ m$^{-1}$ (with $d \approx 87\,\xi$ for instance) new physics could be then still reachable at ILL or ESS!

At non-relativistic energies, the coupling Hamiltonian between each brane is given by (Petit and Sarrazin, 2005; Sarrazin and Petit, 2010 and 2012):

$$H_{cm} = i(1/2)g\mu \begin{pmatrix} 0 & -\vec{\sigma}\cdot(\vec{A}_+ - \vec{A}_-) \\ \vec{\sigma}\cdot(\vec{A}_+ - \vec{A}_-) & 0 \end{pmatrix} \tag{2}$$

where $\mu$ is the magnetic moment of the particle and $A_+$ and $A_-$ the magnetic vector potentials in each brane. It can be shown that for a bulk containing at least two parallel 3-branes hidden to each other, matter swapping between these two worlds must occur according to Eq. (2). The particle must be endowed with a magnetic moment. This swapping effect between two neighboring 3-branes is triggered by using suitable magnetic vector potentials. More important, this effect could be detected with present day technology which opens the door to a possible experimental confirmation of the braneworld hypothesis. For charged particles, the swapping is possible though a bit more difficult to achieve. As a consequence, for the sake of simplicity and in order to be able to distinguish the swapping effect from other kinds of predicted phenomena, the use of neutrons must be considered for a prospective experiment. Actually, cold neutron experiments related to the neutron disappearance are fundamental since they could allow to quantify or to distinguish among the different predicted phenomenologies (Ban, al., 2007; Sarrazin and Petit, 2012). It is also relevant to consider astrophysical magnetic vector potentials, which can lead to such a matter swapping. Indeed, the basic argument is that the astrophysical vector potentials are considerably larger than any other counterpart generated in a laboratory (Sarrazin et al., 2012). A possible consequence for free neutrons would be then high frequency and small amplitude oscillations of the matter swapping probability between the two branes. Cold neutrons striking a wall would therefore have a non-zero probability $p$ to escape from our brane toward the hidden brane during the collision, with $p = 2g^2\mu^2|\vec{A}_+ - \vec{A}_-|^2/(V_+ - V_-)^2$, where $V_\pm$ is the gravitational potential energy of the particle in each brane. Such a process would be perceived as a neutron disappearance (i.e. a violation of baryon number conservation) from the point of view of an observer located in our brane. Some experimental device designs can be considered to measure $p$ and have still to be tested.

## 2. Preliminary experimental works

In a recent paper (Sarrazin et al., 2012), a first upper limit on the exotic disappearance probability $p$ was given from previous performed experiments ($p < 7 \times 10^{-6}$ at 95% C.L.). Some of these data came from experiments on ultracold neutrons at ILL. In this first concept, ultracold neutrons are stored in a vessel, with a mean wall collision rate $\gamma$ which is typically in the range from 1 Hz to 100 Hz. The number of stored neutrons decays by following a nearly exponential law with a decay time $\tau_{st}$. This storage time $\tau_{st}$ is measured by counting the remaining neutrons after a storage period of variable duration. The inverse of the storage time is the sum of the neutron beta decay rate and the loss rate $\gamma p$ due to wall collisions and which corresponds to a disappearance into the other brane.

## 3. A neutron-shining-through-a-wall experiment

Let us suggest a new sensitive experiment (Sarrazin et al., 2013). Cold neutrons striking a wall have a probability $p$ to escape from our brane toward the hidden brane during the collision. Thus one can build a cold neutron source in the hidden brane. The hidden neutrons can then bypass the wall thanks to the hidden brane. It can then be shown that there is a nonzero probability (equal to $p$) to detect such neutrons arising from the other brane. As a consequence, a neutron which strikes a wall gets a probability about $p^2$ to be detected behind the wall, bypassing it through another brane. This is a "cold-neutron-shining-through-a-wall" experiment. The experiment rests on a simple device mainly constituted by a wall which blocks neutrons and a low-noise neutron detector. The device must be supplied with an intense cold neutron beam, which enhances the expected signal. Three main parts must be considered:

- The wall
  It is a multilayer passive device made of boron carbide (to absorb neutrons) and lead (to absorb secondary radiations due to neutron capture). According to the theory, a weak ratio of neutrons can escape in another brane. Various wall thickness can be considered to discriminate with an unexpected solid state or nuclear effect.
- The low-noise neutron detector
  For example, the LARN of the University of Namur holds a low-noise chamber (Genard et al., 2010), which has been developed to obtain very clean cross-section measurements of some nuclear reactions relevant for astrophysics. A lead castle shielding device is used to reduce the natural radioactivity and an active shielding is supplied by an anticoincidence system between a scintillator and the detector. The setup allows a strong reduction of the background noise providing an improved sensitivity. It can be easily associated with a relevant [3]He neutron detector.

- The cold neutron beam
  As a reference, let us consider the PF1B beam available at ILL facility (Abele et al., 2008). It is a cold neutron beam with mean energies about 4-5 meV (i.e. $\lambda \approx 4.0$-$4.5$ Å). The neutron particle flux $\varphi_{in}$ is about $8 \times 10^{11}$ s$^{-1}$, with a rectangular cross-section of about $60 \times 200$ mm$^2$.

Roughly speaking, the neutron flux behind the wall would be $\varphi = \varphi_{in} p^2$. Let us consider a test estimate of the constraint on the upper value of $\varphi$, says $\varphi < 4 \times 10^{-4}$ s$^{-1}$. This is a fair value while considering the low-noise efficiency of the device of the LARN with a 14h time acquisition. Using the PF1B source, one can then expect for $p < 2 \times 10^{-8}$. Such a constraint would be 350 times better than the present constraint $p < 7 \times 10^{-6}$ (Sarrazin et al., 2012). The actual means of the LARN and of the ILL (and of the future ESS) allow then quite easily for an experiment with a significant sensitivity, which allows a strong improvement of the constraint on $p$, i.e. on the baryon number conservation in braneworlds models.

## 4. Conclusion

Braneworld is a concept inherited from high energy physics, which needs to be tested. New phenomenological consequences are allowed in many-world models and matter swapping between branes (under magnetic vector potential constraint) is predicted. This allows to test the braneworld hypothesis at low energies with cold neutrons. The effect can be discriminated from other effects and is fully relevant for experiments at ESS or at ILL facilities. Using a cold neutron beam (with 4 Å $< \lambda <$ 9 Å for a relevant cross-section) with a minimal flux $\varphi_0 \sim 10^9$ n/s/cm$^2$ and thanks to a low noise neutron detector, it is possible to develop a neutron-shining-through-a-wall experiment. Predicted constraints are better by up to 2 orders of magnitude compared to existing results using storage experiments.

**Acknowledgements**

The author acknowledges Fabrice Petit, Guillaume Pignol, Valery V. Nesvizhevsky and Guy Terwagne for useful discussions or comments.